# FODT: Fast, Online, Distributed and Temporary Failure Recovery Approach for MEC

Xin Yuan, Ning Li, Jose Fernan Martinez

***Abstract*-Mobile edge computing (MEC) can reduce the latency of cloud computing successfully. However, the edge server may fail due to the hardware of software issues. When the edge server failure happens, the users who offload tasks to this server will be affected. How to recover the services for these affected users quickly and effectively is challenging. Moreover, considering that the server failure is continuous and temporary, and the failed server can be repaired, the previous works cannot handle this problem effectively. Therefore, in this paper, we propose the fast, online, distributed, and temporary failure recovery algorithm (FODT) for MEC. In FODT, when edge sever failure happens, only the affected APs recalculate their user-server allocation strategies and the other access points (APs) do not change their strategies. For the affected APs, the strategies before server failure are reused to reduce complexity and latency. When the failed server is repaired, the influenced APs reuse the strategies before server failure to offload task to this server. Based on this approach, the FODT can achieve better performance than previous works, includes higher failure recovery efficiency and lower complexity with acceptable approximate ratio.**

## I. Introduction

### *A. Motivation, Problem and Challenges*

In MEC, the base stations (BS), the communication links, and the edge servers may suffer failure due to the unexpected link failure, software bug, or hardware problems with high probability [1][2]. For instance, a significant fraction of connection failures is as large as 45% among 5 million users using WiFi networks in the urban area [3]. Once the edge services are interrupted, the tasks of mobile users need to be calculated locally or offloaded to the remoted cloud with high latency and resource consumption. Previous studies have investigated how to optimally place the edge servers on BSs to minimize the task calculation delay when there are some servers fail [1][2]. However, the Failure Recovery Issue (FRI) has been ignored by previous works.

The FRI includes two aspects: first, when edge server failure occurs, allocating the affected mobile users to offload tasks to the alternative edge servers (i.e., user-server allocation strategy); second, when the failed servers go back online, rescheduling the user-server allocation strategy. These two aspects are not isolated. The strategies in the first aspect can affect the complexity of the second part seriously. For instance, if all the user-server allocation strategies are recalculated when the server failure happens to achieve low latency for all users, once the failed servers have been repaired, all the user-server allocation strategies need to be recalculated again in the second aspect. This means that the user-server allocation strategies, which are always complex and time-consuming, need to be calculated twice in FRI. Therefore, when designing the failure recovery strategy, these two aspects should be optimized jointly, which has not been investigated in previous works. Thus, the problem needs to be solved in this paper can be summarized as *when server failure happens, how to design the optimal user-server allocation strategy to minimize the latency in the two aspects of FRI.* This is critical to guarantee the low latency and high efficiency of the MEC which may suffer from edge server failure with high probability.

### *B. Disadvantages of Prior works*

The user-server allocation algorithms in edge server deployment (both the algorithms with robustness [1][2] and without robustness [4][5][6]) and the edge server selection algorithms [7][8][9] are related to the failure recovery issues. The purpose of server deployment is to find $k$ optimal locations in $s$ network access points (APs) or base stations (BSs), where $k < s$, and to deploy the edge servers to achieve optimal network performance, such as minimum latency and deployment cost, maximum network throughput, high robustness, etc. When the servers are deployed on the optimal APs or BSs, the other APs or BSs will choose to offload tasks to edge servers under different optimal purposes, which is known as the user-server allocation. The purpose of server selection is to choose the optimal edge server for each mobile user to offload their calculation tasks with different optimization objectives, such as minimizing the average delay [10][11], maximizing the resources utilization efficiency [12][13], minimizing energy consumption [14][15], etc. During server selection, there is a common assumption that each AP or BS associates with an edge server and each mobile user can connect to more than one edge server through wireless manner, which cannot always be satisfied in practice due to the budget and deployment cost, especially for the large-scale MEC in urban area.

However, the user-server allocation algorithms in previous works cannot solve the failure recovery issues effectively due to the following reasons. First and foremost, the server failure is unexpected and continuous, and the same edge server could fail muti-times. This means that the failure recovery is an online problem. Thus, the user-server allocation strategies in previous works, which are already complex, need to be executed frequently when solving FRI. Moreover, since there may be tens of thousands of APs in a metropolitan area, the complex and latency may be serious. Secondly, the second aspect of FRI is ignored in these algorithms. Based on the user-server allocation algorithms in previous works, when edge server failure happens, all the user-server allocation strategies are recalculated. Then once the failed servers go back online, all the user-server allocation strategies need to be recalculated again in the second aspects of FRI. This implies that the user-server allocation strategies, which is always complex and time-consuming, need to be calculated twice. Moreover, considering that the server failure is always temporary (few minutes, hours, or days), the user-server allocation algorithms will be executed frequently. Therefore, applying previous user-server allocation algorithms into FRI will cause serious performance deterioration.

*C. Proposed Approach*

Considering the disadvantages of previous woks, in this paper, we propose the Fast, Online, Distributed, and Temporary failure recovery algorithm (FODT) for addressing FRI in MEC. In FODT, when the edge server failure happens, the direct-affected APs will choose the optimal edge server from its neighbor servers and find the optimal routing to this server with minimum latency. The APs whose servers are not failure will not change their user-server allocation strategies. Moreover, when server failure happens, for achieving minimum recovery latency, first, the direct-affected APs use the reverse routing during the coverage of failed servers to transmit task to the edge of neighbor servers' coverage; then, the edge AP of the failed servers chooses one accessing AP from its neighbor servers' coverage and sends tasks to this AP; finally, this accessing AP will offload the tasks to its server through its optimal routing. When the failed server goes back online, the directed-affected APs will offload task to this server based on the previous user-server allocation strategy. The other APs do not change their strategies. Based on this online and approximate approach, the FODT can achieve better performance than previous research. Moreover, in this paper, we also investigate the properties of FODT, including the approximate ratio, the complexity, and the robustness. The FODT has higher failure recovery efficiency and lower complexity than previous works with acceptable approximate ratio.

## II. System Model and Problem Statement

*A. Network model*

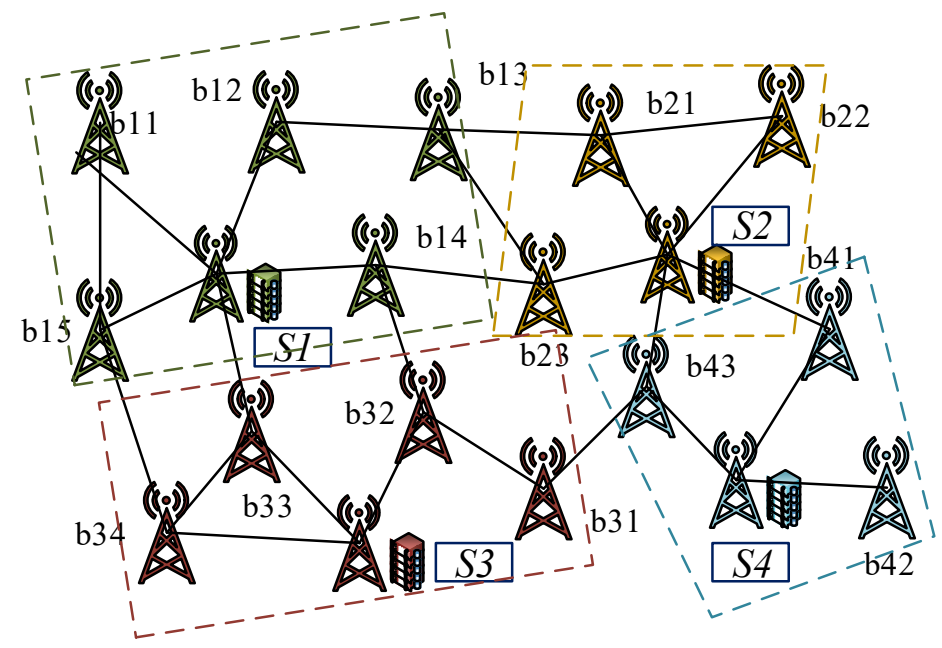

Fig.1. Network model

In FODT, we consider a typical edge computing network where there are $M$ access points (AP), e.g., WiFi, BS, mobile vehicles, etc. The APs are interconnected by the cellular network [16], which provides efficient and flexible communications including data and control information among APs. Denote the sets of $M$ APs and $L$ edge servers as $N = \{N_1, N_2, \dots, N_M\}$ and $S = \{S_1, S_2, \dots, S_L\}$, respectively. We will investigate the scenarios that the servers are heterogeneous in terms of capabilities. The edge servers have already been deployed on $L$ APs based on the previous edge server placement algorithms, such as [1-2, 4-6]. In practice, the number of APs can be huge. For instance, according to the trace data from Shanghai Telecom, there are 389 base stations in only 43 square kilometers of central urban area of Shanghai. Thus, in our model, $L < M$ is assumed, based on the network models that used in previous works and practical, it is reasonable. Due to the finite computing capacity and limited number of edge servers, some computing requests may not be served locally. Besides the locally deployed edge server, the user request can be offloaded to other nearby edge servers or the remote cloud. To serve the request with low latency, the request should not be offloaded to distant servers [17]. Consequently, as shown in Fig.1, for each edge server, it can provide services for multiple APs based on optimal user-server allocation algorithms. Moreover, since the request can be offloaded to nearby edge servers through more than one hop transmission by APs, we give two definitions as follows to quantify this model.

**Definition 1.** *Coverage*. The coverage of edge server $s$ is defined as the set of APs that served by $s$, denoted as $C_s$.

As shown in Fig.1, since the $b_{11}$, $b_{12}$, $b_{13}$, $b_{14}$ and $b_{15}$ are all offloading their tasks to server $s_1$, $C_{s_1} = \{b_{11}, b_{12}, b_{13}, b_{14}, b_{15}\}$.

**Definition 2.** *Coverage depth*. The coverage depth of server $s$, i.e., $d_s$, is defined as the maximum number of hops that the AP in $C_s$ can offload tasks to server $s$.

For instance, as depicted in Fig.1, the number of hops from $b_{13}$ to $s_1$ is 2, which is the maximum values during $b_{11}$, $b_{12}$, $b_{13}$, $b_{14}$ and $b_{15}$, then $d_{s_1} = 2$.

The delay in MEC includes transmission delay, propagation delay, and processing delay. Since the propagation delay is usually negligible compared to the transmission and processing delay, we only calculate the transmission delay and processing delay in this model. The notations used in this model are presented in Table 1.

*B. Transmission delay model*

The average task queue length in AP $j$ at slot $z$, denoted as $\bar{Q}_{BS}^{j-z}$, can be calculated as:

$$\bar{Q}_{BS}^{j-z} = \frac{1}{t_z}\{t_z(T_U^j + T_{ne}^j) - t_z f_{BS}^j\} + \bar{Q}_{BS}^{j-(z-1)} = T_U^j + T_{ne}^j - f_{BS}^j + \bar{Q}_{BS}^{j-(t-1)} \quad (1)$$

where $T_U^j$ is the tasks that transmitted from mobile users to AP $j$, $T_{ne}^j$ is the tasks that transmitted from neighbor APs of $j$, $f_{BS}^j$ is the probability of AP $j$ on dealing with task transferring, measured in units of million bits per second, i.e., Mbps, $t_z$ is the duration of slot $z$.

As a result, $T_U^j$ can be calculated as:

$$T_U^j = \sum_{i=1}^{N} \alpha_i^j \lambda_u^i \quad (2)$$

where $N$ is the number of mobile users in network, $\lambda_u^i$ is the requirements transmission rate of mobile user $i$; $\alpha_i^j$ represents the proportion of tasks of mobile user $i$ that offloaded to AP $j$. Moreover, considering the property of task partition, the whole task can be divided into $w$ discrete values, i.e., $\alpha_i^j \in \{\alpha_1, \alpha_2, \dots, \alpha_w\}$ with constructs $\alpha_i \in [0,1]$ and $\sum_{i=1}^{k} \alpha_i = 1$. The task offloading issues are investigated widely and they are not the main research contents of this paper, which can be calculated based on previous algorithms, such as [1], [2].

Moreover, $T_{ne}^j$ can be calculated as:

$$T_{ne}^j = \sum_{i=1}^{M} \beta_i^j \lambda_B^i p_i^j \quad (3)$$

where $M$ is the total number of APs in network; $\lambda_B^i$ is the requirements transmission rate of AP $i$; $\beta_i^j$ represents whether AP $i$ choose AP $j$ as forwarder, i.e., $\beta_i^j = 1$ means that AP $i$ chooses AP $j$ as forwarder; otherwise $\beta_i^j = 0$; $p_i^j$ represents the proportion of tasks of AP $i$ that transferred to AP $j$. Moreover, considering the property of task partition, the whole task can be divided into $n$ discrete values, i.e., $p_i^j \in \{p_1, p_2, \dots, p_n\}$ with constructs $p_i \in [0,1]$ and $\sum_{i=1}^{n} p_i = 1$. Thus, $\bar{Q}_{BS}^j$ can be expressed as:

$$\bar{Q}_{BS}^j(\beta^j, p^j) = \sum_{i=1}^{N} \alpha_i^j \lambda_u^i + \sum_{i=1}^{M} \beta_i^j \lambda_B^i p_i^j - f_{BS}^j + \bar{Q}_{BS}^{j-(z-1)} \quad (4)$$

Based on (4), the transmission delay in AP $j$ can be computed as:

$$D_t^j(\beta^j, p^j) = \frac{\bar{Q}_{BS}^j(\beta^j, p^j)}{f_{BS}^j} = \frac{\sum_{i=1}^{N}\alpha_i^j\lambda_u^i + \sum_{i=1}^{M}\beta_i^j\lambda_B^i p_i^j - f_{BS}^j + \bar{Q}_{BS}^{j-(z-1)}}{f_{BS}^j} \quad (5)$$

where $\beta^j = \{\beta_1^j, \beta_2^j, \dots, \beta_M^j\}$ and $p^j = \{p_1^j, p_2^j, \dots, p_M^j\}$, respectively.

*C. Processing Delay Model*

The average queue length in edge server $l$ at slot $z$, denoted as $\bar{Q}_{es}^{l-z}$, can be calculated as:

$$\bar{Q}_{es}^{l-z} = \frac{1}{t_z}\{t_z \cdot T_{bs}^l - t_z f_{es}^l\} + \bar{Q}_{es}^{l-(z-1)} = T_{bs}^l - f_{es}^l + \bar{Q}_{es}^{l-(t-1)} \quad (6)$$

where $T_{bs}^l$ is the task transmitted ratio from APs to edge server $l$; $f_{es}^l$ is the capability of edge server $l$ on processing tasks, measured in units of Billion bits per second, i.e., Bbps. In (6), $T_{bs}^l$ can be calculated as:

$$T_{bs}^l = \sum_{i=1}^{M}\gamma_i^l\lambda_B^i \quad (7)$$

where $\gamma_i^l = 1$ represents that AP $i$ offloads its tasks to edge server $l$; otherwise $\gamma_i^l = 0$. Assuming that the number of edge servers is $L$, then we have $\sum_{l=1}^{L}\gamma_i^l = 1$.

Thus, $\bar{Q}_{es}^l$ can be expressed as:

$$\bar{Q}_{es}^l(\gamma^l) = \sum_{i=1}^{M}\gamma_i^l\lambda_B^i - f_{es}^l + \bar{Q}_{es}^{l-(t-1)} \quad (8)$$

The processing delay in edge server $l$ can be calculated as:

$$D_p^l(\gamma^l) = \frac{\bar{Q}_{es}^l(\gamma^l)}{f_{es}^l} = \frac{\sum_{i=1}^{M}\gamma_i^l\lambda_B^i - f_{es}^l + \bar{Q}_{es}^{l-(t-1)}}{f_{es}^l} \quad (9)$$

where $\gamma^l = \{\gamma_1^l, \gamma_2^l, \dots, \gamma_M^l\}$.

Based on (5) and (9), if AP $j$ is in the coverage of server $l$ and wants to offload tasks to it, then the delay (including processing delay and transmission) can be calculated as:

$$D_j^l(\beta^j, p^j, \gamma^l) = D_t^j(\beta^j, p^j) + D_p^l(\gamma^l) = \frac{\bar{Q}_{es}^l(\gamma^l)}{f_{es}^l} + \frac{\bar{Q}_{BS}^j(\beta^j, p^j)}{f_{BS}^j} \quad (10)$$

*D. Problem statement*

According to the transmission delay model and processing delay model, the purpose of failure recovery is to decide the user-server allocation strategies and task offloading strategies, i.e., $\beta$, $p$, and $\gamma$, for the two aspects of failure recovery issue to minimize the task transmission and processing delay. When server $s$ fails, it will trigger the calculation of user-server allocation, and the delay is:

$$D_F = \sum_{j=1}^{M}\sum_{l=1}^{L} D_j^l(\boldsymbol{\beta_F}, \boldsymbol{p_F}, \boldsymbol{\gamma_F}) \quad (11)$$

where $\boldsymbol{\beta_F} = \{\beta_F^1, \beta_F^2, \dots, \beta_F^M\}$ and $\boldsymbol{p_F} = \{\rho_F^1, \rho_F^2, \dots, \rho_F^M\}$ are the task offloading strategies when server fails, and $\boldsymbol{\gamma_F} = \{\gamma_F^1, \gamma_F^2, \dots, \gamma_F^M\}$ is the user-server allocation strategy. When server $s$ is repaired, the user-server allocation strategy needs to be computed again. Then the delay in this aspect can be expressed as:

$$D_R = \sum_{j=1}^{M}\sum_{l=1}^{L} D_j^l(\boldsymbol{\beta_R}, \boldsymbol{p_R}, \boldsymbol{\gamma_R}) \quad (12)$$

where $\boldsymbol{\beta_R} = \{\beta_R^1, \beta_R^2, \dots, \beta_R^M\}$ and $\boldsymbol{p_R} = \{\rho_R^1, \rho_R^2, \dots, \rho_R^M\}$ are the task offloading strategies when the failed server goes back online, and $\boldsymbol{\gamma_R} = \{\gamma_R^1, \gamma_R^2, \dots, \gamma_R^M\}$ is the user-server allocation strategy.

Then, the problem to be solved in this paper can be denoted as:

$$\text{(P0)} \quad \min \sum_{l=1}^{L}\rho(D_F + D_R) \quad (13)$$

$$s.t. \quad \beta_i^j \in \{0,1\},\ \forall i, j \in [0\ M] \quad (13.a)$$

$$p_i^j \in [0,1]\ \&\ \sum_{i=1}^{n} p_i = 1,\ \forall i, j \in [0\ M] \quad (13.b)$$

$$\sum_{l=1}^{L}\gamma_i^l = 1 \quad (13.c)$$

$$f_{BS}^j \le f_{BS}^{max} \quad (13.d)$$

$$f_{ES}^j \le f_{ES}^{max} \quad (13.e)$$

In (13), $L$ is the number of servers that deployed in the whole network, $\rho \in [0,1]$ represents the probability of edge server failure. Constraint (a) guarantees that each AP can offload task to edge server directly or assisted by other APs; constraint (b) means that all the tasks in AP should be offloaded to edge server; constraint (c) guarantees that for each AP, its task can only offload to one edge server; constraint (d) and (e) guarantee that the capability of AP and edge server cannot exceed the maximum value.

## III. The details of FODT

*A. P1: An approximate model of P0*

When one edge server fails, the APs that served by the failed server need to offload their calculation task to their neighbor edge servers to minimize the calculation latency, which will affect the task processing latency of APs that served by the neighbor edge servers. Therefore, based on the different status of edge servers and APs when the server failure happens, we divide the edge servers and APs into two different categories: 1) the direct-affected AP, 2) the indirect-affected AP and ES, which are defined in Definition 3 and Definition 4, respectively. The direct-affected AP is the AP that covered by the failed edge server. For instance, as shown in Fig.1, when server $s_1$ fails, the $b_{11}$, $b_{12}$, $b_{13}$, $b_{14}$ and $b_{15}$ who have selected $s_1$ as their task processing server, are the direct-affected APs. When server $s$ fails, the direct-affected APs will re-calculate the user-server allocation strategies and the task offloading strategies, which will affect other APs. For instance, if $b_{14}$ selects server $s_3$ as its new task processing server, then the processing delay of all the APs that covered by $s_3$ will be influenced. The indirect-affected ESs are the servers which handle the task offloading requirements from the direct-affected APs, and the APs that served by these servers are the indirect-affected APs. For instance, as displayed in Fig.1, the $b_{21}$~$b_{23}$ and $b_{31}$~$b_{34}$ are all indirect-affected APs.

In the approximate model, for the unaffected APs and ESs, their user-server allocation strategies and the task offloading strategies are not changed. Only the strategies of the indirect-affected APs and the direct-affected APs need to be recalculated under the purpose of calculation latency minimization. Thus, assuming that there are $s'$ servers failed and the number of APs that covered by these $s'$ servers is $h$, the number of indirect-affected ESs is $x$ and the number of APs that served by these $x$ servers is $y$, then the number of APs declines to $h+y$ and the number of servers reduces to $s'+x$ compared with P0. Thus, when server failure happens, the delay caused by user-server allocation strategy calculation is:

$$D_F' = \sum_{j=1}^{h+y}\sum_{l=1}^{s'+x} D_j^l(\boldsymbol{\beta_F}, \boldsymbol{p_F}, \boldsymbol{\gamma_F}) \quad (14)$$

The delay resulted from the failed server goes back online is:

$$D_R' = \sum_{j=1}^{h+y}\sum_{l=1}^{s'+x} D_j^l(\boldsymbol{\beta_R}, \boldsymbol{p_R}, \boldsymbol{\gamma_R}) \quad (15)$$

According to (14) and (15), the approximate model of P0, i.e., P1, can be expressed as:

$$\text{(P1)} \quad \min \sum_{l=1}^{s'+x}\rho(D_F' + D_R') \quad (16)$$

$$s.t. \quad (14.a) - (14.e)$$

Compared to P0, the complexity of P1 has been decreased when dealing with failure recovery issues. However, from (13) and (16), we can find that actually P1 is the sub-problem of P0. The reason is as follows. The purpose of P0 is to find the optimal user-server allocation strategies between all $M -$ $S$ APs and $S$ edge servers; the purpose of P1 is to find the optimal user-server allocation strategies between $h+y$ direct-affected and indirect-affected APs and $s'+x$ failed and indirect-affected servers. When considering the server failure probability, then $s'=\rho S$, $h=l\rho S$, and $y=lx$. Moreover, $x$ is related to $s'$ and $s'$ is associated with $\rho$. Hence, we have $x=f(\rho)$. Then the purpose of P1 is to find the optimal user-server allocation strategies between $l\rho S+lf(\rho)$ APs and $\rho S+f(\rho)$ servers. Therefore, the algorithm for solving P0 and P1 are the same, the only difference between them is that the scale of the input data in P1 is much smaller than that in P0 when the server failure probability is not high. Therefore, P1 is also NP-hard. However, based on the approximate model in P1, it is much more effective on dealing with the online problem than P0. As a result, in the following of this section, we will propose FODT which is the approximate solution of P1.

*B. FODT: the approximate solution of P1*

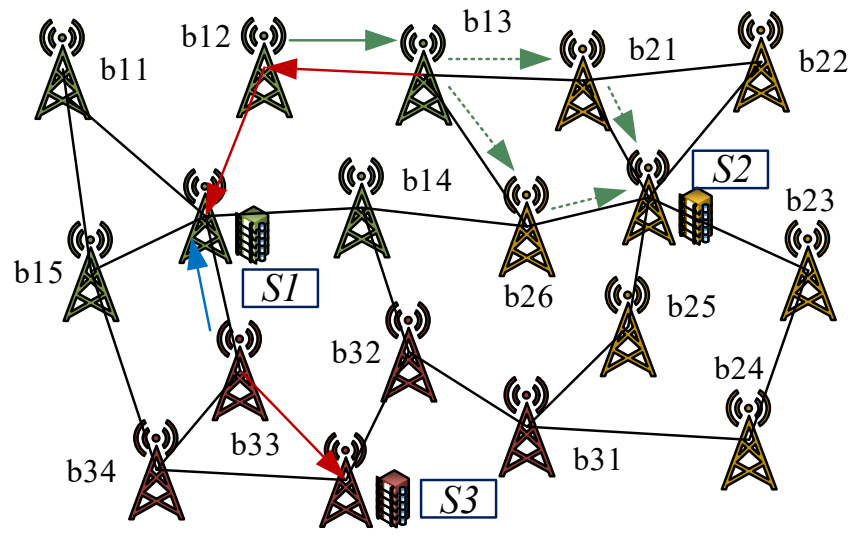

Fig.2. The principle of FODT

As depicted in Fig.2, before edge server failure, the optimal user-server allocation strategies have been calculated based on previous server placement algorithms, such as [1-2, 4-6]. Therefore, within each server's coverage, the routing of each AP to this server is already optimized. For instance, in Fig.1, in the coverage of edge server $s_1$, all the BSs within the green circle have the optimal routing to $s_1$. This is the same as other edge servers. Once edge server $s_1$ fails, $b_{11}$, $b_{12}$, $b_{13}$, $b_{14}$, and $b_{15}$ will launch the failure recovery process, which includes finding an optimal edge server and offloading tasks to this server for minimizing the task calculation latency. When server $s_1$ goes back online, $b_{11}$, $b_{12}$, $b_{13}$, $b_{14}$, and $b_{15}$ use the former user-server allocation strategy before $s_1$ fails to minimize the calculation and processing latency. The FODT includes three steps.

*Step1*. Routing reserving in the coverage of the failed servers. As shown in Fig.2, before server $s_1$ fails, the task offloading routing from $b_{13}$ to $s_1$ is $b_{13}\rightarrow b_{12}\rightarrow s_1$. Then, when $s_1$ fails, if $b_{12}$ connects to one neighbor server, it will offload tasks to this server directly; otherwise, if $b_{12}$ does not connect to any neighbor server straight, it reverses the task offloading routing from $b_{12}$ to $b_{13}$, i.e., the task offloading routing becomes $b_{12}\rightarrow b_{13}$. By this reversed routing, the task in $b_{12}$ will be brought to the edge of the coverage of $s_1$. The other APs in the coverage of $s_1$ use the same routing reversing approach to transmit their tasks to the edge of the coverage of $s_1$. Based on the routing reversing approach, on the one hand, since the latency of routing $b_{13}\rightarrow b_{12}\rightarrow s_1$ is optimal, the latency of the reversed routing $b_{12}\rightarrow b_{13}$ can be guaranteed. On the other hand, by using the reversed routing, we do not need to start a route construction process which is time-consuming. So, by the routing reversing, the latency of failure recovery can be guaranteed.

*Step2*. Finding the optimal accessing APs in neighbor servers' coverage. As illustrated in Fig.2, when the tasks in $b_{12}$ is transmitted to $b_{13}$ which is the edge APs of the coverage of $s_1$, the next step is to decide which neighbor server is selected to process these tasks and what is the optimal routing to this server. In Fig.2, the $b_{13}$ connects to two APs except for the APs covered by $s_1$, which are $b_{21}$, and $b_{26}$. Moreover, since the user-server allocation strategies in the indirect-affected servers are not changed, the routes from $b_{21}$ and $b_{26}$ to $s_2$ are known in advance. The average delay of $b_{21}$ and $b_{26}$ is $D_{21}$ and $D_{26}$, respectively, which includes both the transmission delay and calculation delay, and is known based on previous server placement algorithms, too. In FODT, we regard the $b_{21}$ and $b_{26}$ as the accessing APs of server $s_2$. Therefore, in this step, the $b_{13}$ only needs to decide which AP is selected as the accessing AP. When the accessing AP is decided, the related server is determined, too. The principle of AP selection is to choose the AP whose delay is the minimum one. For instance, for $b_{13}$, it has two alternative routes to $s_2$, i.e., $b_{13}\rightarrow b_{21}\rightarrow s_2$ and $b_{13}\rightarrow b_{26}\rightarrow s_2$; thus, if the latency (including the transmission delay and processing delay in $s_2$) in $b_{21}$ is less than that in $b_{26}$, the $b_{13}\rightarrow b_{21}\rightarrow s_2$ is chosen; otherwise, $b_{13}\rightarrow b_{26}\rightarrow s_2$ is adopted. Once the optimal edge server and routing is selected, the direct-affected APs can transmit tasks based on this approach.

*Step3*. Rescheduling the user-server allocation strategies when the failed server works normal again. When the function of the failed server goes back online, the direct-affected APs will reuse the former user-server allocation strategies before server failure happens. For instance, as shown in Fig.2, when the failed server $s_1$ is repaired, $b_{13}$ will choose $s_1$ as its server again and use the former routing, i.e., $b_{13}\rightarrow b_{12}\rightarrow s_1$, to offload tasks to $s_1$. And the user-server allocation strategies of the indirect-affected APs in $s_2$ do not need to re-calculate when $s_1$ works normal again. Based on this approach, the latency for calculating the user-server allocation strategies when the failed servers work as usual is small.

## IV. The properties of FODT

In this section, we investigate the properties of FODT, including the approximate ratio, the complexity, and the robustness.

**Corollary 1**. The approximate ratio of FODT is $\frac{1+\rho(\omega-1)}{\tau}$, where $\rho$ is the probability of server failure, $\omega$ is the times of the delay of the affected APs using FODT to the original delay of the APs that served by the selected edge servers, $\tau$ is the times of the delay of the affected APs applying FODT to the original delay of the APs that served by the selected edge servers.

*Proof.* Assuming that the number of APs is $N$, the number of edge servers that deployed on APs is $S$, and the number of failed edge servers is $s'$. Moreover, recalling that the optimal solution of P0 means that when there are $s'$ servers' failure, all the APs will recalculate the optimal user-server allocation strategy to make the total task calculation delay minimum, denoted by $D^*_{P0}$.

For FODT, the delay of the unaffected APs will not be influenced by the server failure. The APs that served by the failure servers will offload their tasks to the nearby edge servers through FODT. Then, we assume that the average neighbor servers for each edge server is $k$, and the number of the indirect-affected servers is $x$. Thus, the delay for the proposed approximate algorithm can be expressed as:

$$D_{P1}^* = D_{S-x-s'}^* + D_{x+s'}^* \tag{17}$$

However, in (17), $D_{S-x-s'}^*$ is not affected by the server failure and the same as the value that before edge server failure. $D_{x+s'}^*$ is modified compared with the delay before edge server failure according FODT. In FODT, the user-server allocation strategies of the indirect-affected APs are not changed. Only the direct-affected APs need to reselect the optimal edge server and find the optimal routing to this server. Thus, $D_{x+s'}^*$ can be divided into two different parts logically: 1) the original delay of the indirect-affected APs that served by $x$ indirect-affected edge servers, i.e., the delay of these $x$ servers before the direct-affected APs offload their task to them, denoted by $D_x^*$, and 2) the increased delay of the indirect-affected APs that caused by the task offloading of the direct-affected APs and the delay of direct-affected APs, denoted by $D_{s'}'$. Thus, the $D_{S-x-s'}^*$ can be calculated by:

$$D_{S-x-s'}^* = \bar{l}\bar{D}_{op}(S-x-s') \tag{18}$$

where $\bar{l}$ is the average number of APs that one edge server covered and can be computed as $\bar{l} = \frac{N-S}{S}$; $\bar{D}_{op}$ is the average delay of each AP; $s' = \rho S$. Similarly, the $D_{x+s'}^*$ can be calculated as:

$$D_{x-s'}^* = \bar{l}\bar{D}_{op}(x-s') \tag{19}$$

$$D_{s'}' = \bar{l}\bar{D}_{op}' s' \tag{20}$$

Moreover, $D_{s'}'$ is the increased delay by $s'$ servers' failure. And the average number of APs that served by these servers are $s'\bar{l}$. Since each failed server has $k = \frac{x}{s'} > 1$ neighbor servers, for each indirect-affected server, it will increase $\frac{\bar{l}}{k}$ BSs that offload tasks to it and $\frac{\bar{l}}{k} < \bar{l}$. Moreover, since we select the optimal edge server and routing for each affected AP, for each indirect-affected server, the increased delay is $\frac{\bar{l}s'}{x}\bar{D}_{op}'$. For each $\bar{D}_{op}'$, let $\bar{D}_{op}' = \omega\bar{D}_{op}$, where $\omega > 1$. Therefore, the $D_{P1}^*$ can be expressed as:

$$\begin{aligned} D_{P1}^* &= D_{S-x-s'}^* + D_{x+s'}^* \\ &= \bar{l}\bar{D}_{op}(S-x-s') + \bar{l}\bar{D}_{op}(x-s') + D_{s'}' \\ &< \bar{l}\bar{D}_{op}(S-s') + \omega\bar{l}\bar{D}_{op}s' \\ &= \bar{l}\bar{D}_{op}S[1+\rho(\omega-1)] \end{aligned} \tag{21}$$

Moreover, $D_{P0}^*$ can be calculated as:

$$D_{P0}^* = D_S' = \bar{l}'\bar{D}_{op}''(S-s') \tag{22}$$

where $\bar{l}' = \frac{N-S}{S-s'} = \frac{\bar{l}}{1-\rho}$. Let $\bar{D}_{op}'' = \tau\bar{D}_{op}$, and $\tau > 1$. Thus, the approximate ratio $\epsilon$ can be derived as:

$$\epsilon = \frac{D_{P1}^*}{D_{P0}^*} = \frac{\bar{l}\bar{D}_{op}S[1+\rho(\omega-1)]}{\frac{\bar{l}}{1-\rho}\tau\bar{D}_{op}(1-\rho)S} = \frac{1+\rho(\omega-1)}{\tau} \tag{23}$$

Thus, the approximate ratio is $\frac{1+\rho(\omega-1)}{\tau}$. ■

Based on the Corollary 1, we have conclusion as follows.

**Corollary 2**. Given the server failure probability $\rho$, for FODT, the lower boundary of $\omega$ is $1+\frac{\tau-1}{\rho}$, i.e., $\omega > 1+\frac{\tau-1}{\rho}$; and with the increasing of $\rho$, the $\omega$ increases.

*Proof.* From the Corollary 2, we know that the approximate ratio of FODT is $\frac{1+\rho(\omega-1)}{\tau}$. Moreover, considering that the approximate ration is larger than 1, then we have $\frac{1+\rho(\omega-1)}{\tau} > 1$, which equals to:

$$\tau < 1+\rho(\omega-1) \tag{24}$$

$$\omega > 1+\frac{\tau-1}{\rho} \tag{25}$$

For (24), since $\tau < 1+\rho(\omega-1)$, let $\tau = y[1+\rho(\omega-1)]$ and $y < 1$. Then let:

$$\begin{aligned} F(\rho) &= 1+\frac{\tau-1}{\rho} = 1+\frac{y[1+\rho(\omega-1)]-1}{\rho} \\ &= 1+(\omega-1)y-\frac{1-y}{\rho} \end{aligned} \tag{26}$$

Since $\omega > F(\rho)$, we can conclude that with the increasing of $\rho$, $\omega$ increases, too. ■

**Corollary 3**. The complexity of FODT is $O\left(\frac{M^2}{L}\right)$.

*Proof.* In FODT, as presented in Section IV, for each direct-affected AP, when it recovers the services, it will find the routings from this AP to its neighbor edge servers and choose the optimal one, i.e., the edge server and the routing to this server that minimizes the processing delay of this task. Thus, assuming that for each edge server, the average number of its neighbor edge servers is $L^*$. For each coverage of edge server, the average number of covered APs is $\frac{M}{L}$. As a result, if one edge server fails, the direct-affected APs will offload their tasks to the neighbor coverages, as shown in Fig.2. Therefore, the average number of boundary APs that connected with the direct-affected APs in each neighbor coverages can be calculated as: $\frac{2\pi}{L^*}\cdot\frac{M}{L} = \frac{2\pi M}{L^* L}$. So, for each direct-affected APs, the average number of pairs of edge servers and routing to these servers is $\frac{2\pi M}{L^* L}\cdot L^* = \frac{2\pi M}{L}$. Because the FODT is distributed and each AP decides its own strategies independently, then the average complexity of FODT $O\left(\frac{2\pi M}{L}\cdot M\right)$. Since $2\pi$ is constant, it equals to $O\left(\frac{M^2}{L}\right)$. ■

Note that the average complexity of FODT is much smaller than that in P1. For P1, in the distributed scenario, it is NP-hard; in the centralized scenario, it is $O(LM^2)$. Let $M = \theta L$ and $\theta > 1$, then the complexity of FODT is $O(\theta^2 L)$ under distributed scenario compared to the complexity is $O(\theta^2 L^3)$ in P1 under centralized scenario.

**Corollary 4**. For the given latency threshold $D_{th}$, the algorithm can tolerate at most $s_t$ edge server failure, where $s_t \leq \frac{\left(\frac{\varepsilon_l \upsilon^2 f_{bs}^{min} D_{th}}{T^{min}} - \frac{Q_{min}}{T^{min}}\right)N^2}{M+\left(\frac{\varepsilon_l \upsilon^2 f_{bs}^{min} D_{th}}{T^{min}} - \frac{Q_{min}}{T^{min}}\right)N}$.

*Proof.* Assuming that there are $S$ servers failure, for each un-failed edge server, the average increased number of APs that belongs to the failed edge servers is $\bar{n}$, and the average number of task size of AP is $\bar{T}$. Thus, the average delay of the task can be calculated as:

$$\bar{D} = \frac{Q_{es}'+\bar{n}\bar{T}}{f_{es}'} + y'\frac{Q_{bs}'+\bar{n}\bar{T}}{f_{bs}'} \leq D_{th} \tag{27}$$

Therefore, we have:

$$\begin{aligned} \bar{n} &\leq \frac{1}{\bar{T}}\frac{f_{bs}'f_{es}'D_{th}-(Q_{es}'f_{bs}'+Q_{bs}'y'f_{es}')}{f_{bs}'+y'f_{es}'} \\ &\leq \frac{1}{T^{min}}\frac{f_{bs}^{max}f_{es}^{max}D_{th}-(Q_{es}^{min}f_{bs}^{min}+Q_{bs}^{min}y^{min}f_{es}^{min})}{f_{bs}^{min}+y^{min}f_{es}^{min}} \end{aligned} \tag{28}$$

In (28), $y^{min}$ is the minimum hops from the direct-affected APs to edge servers and $y \geq 1$; thus, we set $y^{min} = 1$. Moreover, let $f_{es}^{max} = \varepsilon_l f_{bs}^{max}$ and $\varepsilon_l > 1$, $f_{es}^{min} = \varepsilon_s f_{bs}^{min}$ and $\varepsilon_s > 1$, $f_{bs}^{max} = \upsilon f_{bs}^{min}$, $Q_{min} = \min\{Q_{es}^{min}, Q_{bs}^{min}\}$, then (28) can be expressed as:

$$\bar{n} \leq \frac{\varepsilon_l \upsilon^2 f_{bs}^{min} D_{th}}{T^{min}} - \frac{Q_{min}}{T^{min}} \tag{29}$$

Moreover, since the total number of APs in the whole network is constant and there are $S$ edge servers fail, then we have:

$$(L-s_t)\left(\frac{M}{N}+\bar{n}\right) = M \tag{30}$$

where $L$ is the total number of edge servers in the network and $M$ is the total number of APs in the network, $\frac{M}{L}$ means the average number of APs in each server's coverage. Based on (30), we have:

$$s_t = \frac{\bar{n}L^2}{M+\bar{n}L} \tag{31}$$

Since $S'(\bar{n}) > 0$, the $S(\bar{n})$ is monotone increasing. Moreover, in practice, $\bar{n}$ is always smaller than $\frac{M}{L}$ because the direct-affected APs of one failure edge servers always do not choose the same target servers. Thus, let $\bar{n} = \lambda\frac{M}{L}$ with $\lambda < 1$, then the (31) can be expressed as:

$$s_t = \frac{\lambda}{1+\lambda}L \tag{32}$$

Thus, this approach can tolerant $\left\lfloor\frac{\lambda}{1+\lambda}L\right\rfloor$ edge server failure. Since $S'(\lambda) > 0$, the $S(\lambda)$ is monotone increasing. Moreover, since $\bar{n} \leq \frac{\varepsilon_l v^2 f_{bs}^{min} D_{th}}{T^{min}} - \frac{Q_{min}}{T^{min}}$ and $S(\lambda)$ is monotone increasing, the upper bound of $s_t$ and $\lambda$ can be calculated as:

$$\lambda \leq \frac{L}{M}\left(\frac{\varepsilon_l v^2 f_{bs}^{min} D_{th}}{T^{min}} - \frac{Q_{min}}{T^{min}}\right) \tag{33}$$

$$s_t \leq \frac{\left(\frac{\varepsilon_l v^2 f_{bs}^{min} D_{th}}{T^{min}} - \frac{Q_{min}}{T^{min}}\right)L^2}{M+\left(\frac{\varepsilon_l v^2 f_{bs}^{min} D_{th}}{T^{min}} - \frac{Q_{min}}{T^{min}}\right)L} \tag{34}$$

**Corollary 5**. If there are *S* edge servers' failure, the proposed algorithm can guarantee the task processing latency is no more and $\bar{D}$, where $\bar{D} \leq \frac{1}{F}(1 + y^{max})(Q + \bar{n}T^{max})$.

*Proof*. Assuming that there are *S* edge servers' failure, then the number of remaining functional servers is $L - S$; thus, the average number of direct-affected APs that joining to the functional servers can be calculated as:

$$\bar{n} = \frac{M}{L-S} - \frac{M}{L} \tag{35}$$

Therefore, the average task processing delay can be calculated as:

$$\bar{D} = \frac{Q_{es}' + \bar{n}\bar{T}}{f_{es}'} + y'\frac{Q_{bs}' + \bar{n}\bar{T}}{f_{bs}'}$$
$$\leq \frac{Q_{es}^{max} + \bar{n}T^{max}}{f_{es}^{min}} + y^{max}\frac{Q_{bs}^{max} + \bar{n}T^{max}}{f_{bs}^{min}} \tag{36}$$

Let $Q = \max\{Q_{es}^{max}, Q_{bs}^{max}\}$ and $F = \min\{f_{es}^{min}, f_{bs}^{min}\}$, then (36) equals to:

$$\bar{D} \leq \frac{Q_{es}^{max} + \bar{n}T^{max}}{f_{es}^{min}} + y^{max}\frac{Q_{bs}^{max} + \bar{n}T^{max}}{f_{bs}^{min}}$$
$$\leq \frac{1}{F}(1 + y^{max})(Q + \bar{n}T^{max}) \tag{37}$$

where $y^{max}$ is the maximum number of hops from direct-affected APs to edge server, which is always small, e.g., 2 or 3 [1]; $T^{max}$ is the maximum size of tasks that transfer from APs to edge servers.

## V. Simulation and Discussion

In this simulation, for considering the number of APs $M$ and the number of deployed servers $L$ simultaneously, we introduce a new parameter into this part, i.e., the server deployment ratio $\mu$ and $\mu = \frac{L}{M}$. The $\mu$ varies from 0.1 to 0.8. the server failure probability varies from 0.1 to 0.8, too. The coverage depth is set to permit task offloading between APs within three hops. Following [17], the request rate $\lambda_i$, the workload of single task $\bar{w}$ and the requested bandwidth of each task $\bar{b}$ are drawn uniformly from the intervals $[3, 5]$ times/s, $[0.5, 1]$ MFLOPs, $[0.5, 1]$ KB/s, respectively. Besides, the available uplink bandwidth $B_i^{up}$ and the available downlink bandwidth $B_i^{down}$ of AP $i$ follow uniform distribution $U[16, 24]$ KB/s. For the edge server settings, the computing capacity of each edge server $s$ complies the uniform distribution with $d_s \sim U[32, 48]$ MFLOPs. Moreover, we choose four benchmarks for performance comparison, i.e., cloud assistant approach, greedy based approach, Robust+ [1], and *kESP-CR* [2].

*1) Performance under different server failure probability*

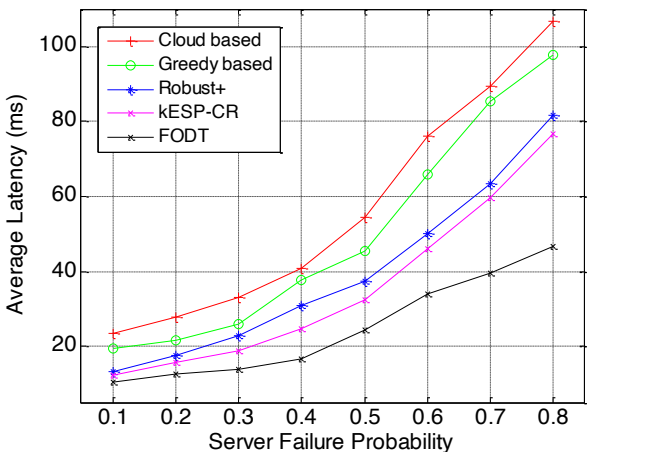

Fig.3. Average latency with different server failure probabilities

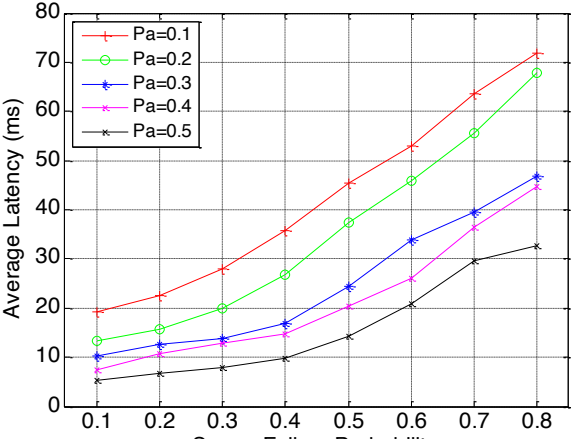

Fig.4. Average latency of FODT with various server failure probabilities

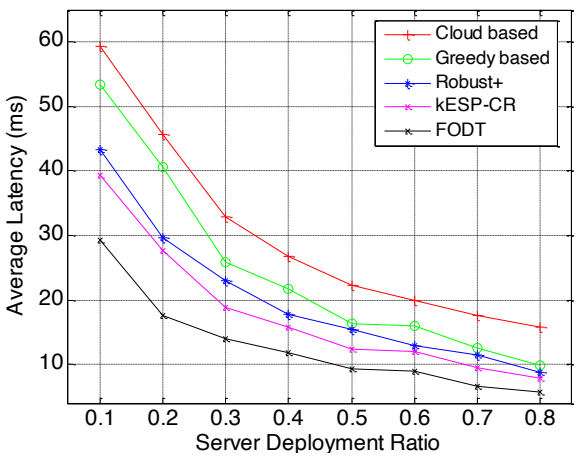

Fig.5. Average latency with different server deployment ratio

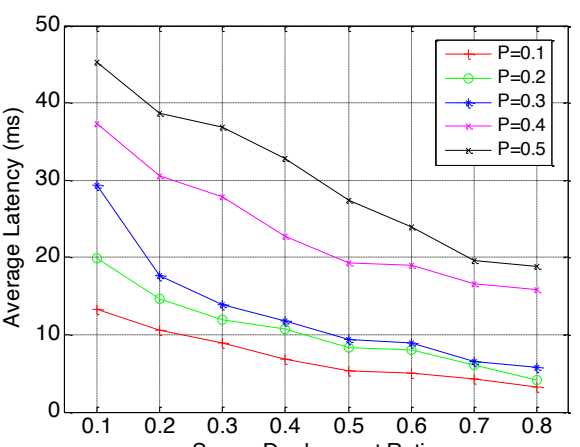

Fig.6. Average latency of FODT under different server deployment ratio

From Fig.3, we can find that with the rising of server failure probability, the average delay in both these five algorithms increase. The performance of FODT is better than that of the other algorithms, and the performance of the cloud assistant approach is the worst. The performance of Robust+ and *k*ESP-CR is similar and much better than that of the greedy-based method. When the server failure probability is small, e.g., fewer than 0.4, the increasing is gently; however, when the server failure probability is large, e.g., bigger than 0.5, the increasing becomes sharply. For instance, when the server failure probability grows from 0.1 to 0.3, the increasing of average latency in FODT is less than 10ms; when the server failure probability increases from 0.5 to 0.7, the latency in FODT gains near 20ms. The same conclusions can be conducted from other algorithms.

In Fig.4, we investigate the average delay of FODT under different server failure probabilities with diverse server deployment ratio, respectively. In this simulation, the server failure probabilities vary from 0.1 to 0.8, and the server deployment ratios are 0.1, 0.2, 0.3, 0.4, and 0.5, respectively. From Fig.4, we can find that with the increasing of server failure probability, the average delay enhances, too. Moreover, the larger the probability is, the obvious of the rising. This is the same as that in Fig.3. Additionally, when the server deployment ratio increases, the average latency decreases under the same server failure probability. Since under the same server failure probability, when the server deployment ratio is large, the ratio of the failed server to the total servers is small, which also means small delay.

*2) Performance under different server deployment ratio*

From Fig.5, we can conclude that with the increasing of server deployment ratio, the average delay of both these five algorithms decreases. Similar to Fig.3, the performance of FODT is the best; the performance of Robust+ and *k*ESP-CR is better than cloud-based approach and greedy-based approach. The reason is the same as that presented in Fig.3. The delay decreasing (when the server deployment ratio is small is faster) than that when the ratio is large. This means for the same server failure probability, the smaller server deployment ratio is, the larger server failure ratio is, which also means large delay. For instance, in FODT, when the server deployment ratio increases from 0.1 to 0.3, the rising of delay is around 15ms; however, when the deployment ratio increases from 0.5 to 0.7, the growth is only 6ms. Thus, in MEC, the more edge servers deployed, the better performance achieves.

The average latency of FODT under different server deployment ratio and various server failure probability is presented in Fig.6. From Fig.6, we can conclude that no matter what the value of server failure probability is, the average delay decreases with the increasing of server deployment ratio. Additionally, for the same server deployment ratio, the latency surges with the enhancing of server failure probability. For instance, when the server deployment ratio is 0.4 and the server failure probability is 0.1, the latency is 7ms; when the server deployment ratio is 0.4 and the server failure probability is 0.4, the delay becomes 23ms. The reason is similar to that in Fig.5.

*3) Performance of convergence when server failure happens*.

From Fig.7, we can find that the convergence time of cloud assistant approach is the minimum one and the convergence time in FODT is close to it. The convergence time in greedy-based approach, Robust+, and *k*ESP-CR is much larger than that in FODT and cloud assistant method. This is because the cloud assistant approach does not need to calculate the user-server allocation strategy. When the server failure happens, the affected APs will offload their tasks to the remote cloud. When these failed server goes back online, the user-server allocation strategies before server failure happens will be reused directly. Thus, its convergence time is the minimum. Moreover, since the FODT considers both the server failure and failed server goes back online, its convergence time is also small. However, since the failed server goes back online does not consider in the other three algorithms, their convergence time is much higher than FODT and the cloud assistant approach. Moreover, with the increasing of network scale, the convergence time of both these five algorithms increase. However, it increases obviously in greedy-based approach, Robust+, and *k*ESP-CR than that in FODT and cloud assistant approach.

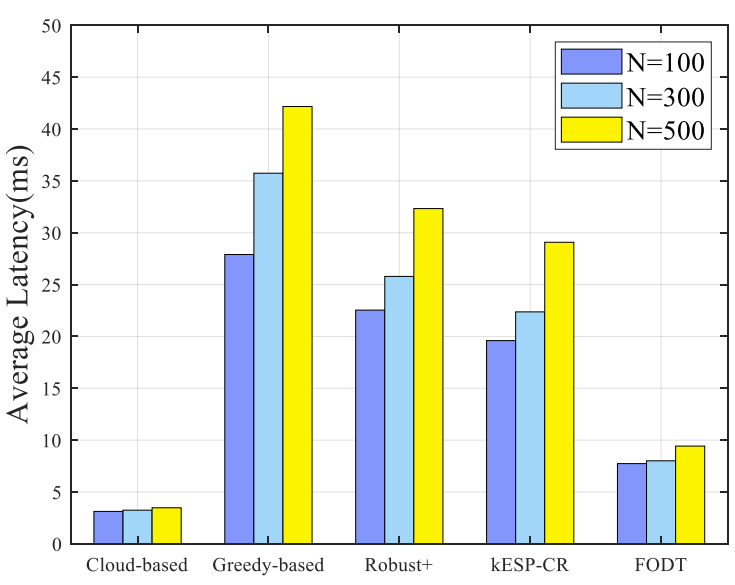

Fig.7. Convergence of different algorithms when server failure

## VII. Conclusions

Considering the disadvantages of previous works on dealing with failure recovery issue, we propose the FODT approach. In FODT, when the edge server failure happens, the direct-affected APs will choose the optimal edge server from their neighbor servers and find the optimal routing to this server. The APs whose servers are not failure will not change their user-server allocation strategies. Moreover, when server failure happens, for achieving the minimum recovery latency, first, the direct-affected APs use the reverse routing during the coverage of the failed servers to transmit the tasks within the edge of the neighbor servers' coverage; then, the edge AP of the failed servers chooses one AP from its neighbor servers' coverage and sends the tasks to this AP; finally, this AP will offload the tasks to its server through its routing to this server. When the failed server goes back online, the directed-affected APs will offload task to the original server based on previous routing; and the other APs do not change their strategies. Based on this online and approximate approach, the FODT can achieve better performance than previous works.

## Acknowledgement

The research leading to the presented results has been undertaken within the National Natural Science Foundation of China under Grant No. 62101159, the Natural Science Foundation of Shandon under Grant No. ZR2021MF055.